\documentclass[jgrga,preprint]{agu2001}

   \usepackage{graphicx}
    \usepackage{amssymb,amsmath}
\authorrunninghead{Breech et al.}

\titlerunninghead{PROTON AND ELECTRON HEATING}

\newcommand{\Csh}{\ensuremath{ C_{\mathrm{sh}} }}
\renewcommand{\d}{\mathrm{d}}

\begin{document}

%

\title{Electron and proton heating by solar wind turbulence
}

%
%

\authors{B.  Breech, \altaffilmark{1}
  W.H.  Matthaeus, \altaffilmark{2}
  S.R.  Cranmer, \altaffilmark{3}
  J.C.  Kasper, \altaffilmark{3}
  and S.  Oughton \altaffilmark{4}}

\altaffiltext{1}{NASA Goddard Space Flight Center}
\altaffiltext{2}{Department of Physics and Astronomy
and Bartol Research Institute, University of Delaware}
\altaffiltext{3}{Harvard-Smithsonian Center for Astrophysics}
\altaffiltext{4}{Department of Mathematics, University of Waikato,
                 Hamilton, New Zealand}

\begin{abstract}
  Previous formulations of heating and transport associated with
  strong magnetohydrodynamic (MHD)
  turbulence are generalized to  incorporate  separate
  internal energy equations for electrons and protons.  Electron heat
  conduction is included.  Energy is supplied by turbulent heating that
  affects both electrons and protons, and is exchanged between them via
  collisions.   Comparison to available Ulysses data shows that a
  reasonable accounting  for the data is provided when
        (i) the energy
  exchange timescale is very long and
        (ii) the deposition of heat due to
  turbulence is divided,  with 60\% going to proton heating and 40\%
  into electron heating.   Heat conduction, determined here by an
  empirical fit, plays a major role in describing the electron data.
\end{abstract}

\begin{article}

\section{Introduction}
\label{sec:intro}

The solar wind displays a highly non-adiabatic temperature profile,
requiring some process(es) to provide additional heat sources.   One
possible, and successful, source of heating comes from
magnetohydrodynamic (MHD) turbulence
present in the solar wind \citep{Coleman68}.   An active MHD turbulent
cascade \citep{MacBrideEA08,MarinoEA08} transfers energy from the
large-scale fluctuations down to small scales where kinetic processes
dissipate the energy as heat.
Previous theories
        \citep{ZhouMatt90a,
                MarschTu93c,
                OughtonMatt95,
                ZankEA96,MattEA99-swh,
                SmithEA01,SmithEA06-pi,
                MattEA04-Hc,BreechEA05,
                IsenbergEA03,
                BreechEA08},
have been able to account
for radial evolution of fluctuation level, correlation scale, cross
helicity and temperature in a specified background solar wind
flow.
These theories
        (with some exceptions
           \citep[e.g.,][]{CranmerEA07})
have focused only on proton temperature,
ignoring heat conduction  and, in effect, assuming that
dissipation of turbulence occurs only through proton kinetic channels.
While neglect of proton heat conduction is justified,
it is not obvious
why all turbulent heating should impact protons alone, nor is it clear
why, or whether, proton-electron energy exchange can be neglected.

Electrons provide additional heating by carrying the bulk of the
solar wind heat flux and through collisions with the
protons.   Turbulence models often neglect electrons in favor of the
protons as the protons help set the scales of interest.  This is
particularly clear in the case of the momentum content and the mass
density of the solar wind plasma.  However the internal energy
content of the electrons is not negligible compared to that of the protons.

For this work, we seek to understand how electrons and protons
conspire to heat the solar wind.   Two possible avenues
immediately open up to explore these issues:  1) an empirically
based approach to compute the heating rates and 2) a modelling
approach based on turbulence theory.   A companion paper
\citep{CranmerEA09} follows the first avenue, e.g.,
examines effects of electrons through a more empirically based
approach.
Here we focus on modelling the heating of the solar wind
through  MHD
turbulence theory.   For the first time, effects of electron heat conduction
are included
in a turbulence transport model that extends beyond the inner
heliosphere.   This provides an important step toward realism and
completeness in the turbulent heating models,
and enables the use of additional observational constraints.

Under the assumptions of a spherically expanding, radially constant
solar wind, the equations for the steady-state temperatures of
electrons, $T_e$, and
protons, $T_p$, may be written as
\begin{equation}
  \label{eq:e-temp-eq}
   \frac{ \d T_e} { \d r} = -\frac{4}{3} \frac{T_e}{r}
                     + \frac{2}{3} \frac{1}{k_B n_e U} Q_e
                     - \frac{T_e - T_p}{U \tau}
                     - \frac{2}{3} \frac{1}{k_B n_e U} \nabla\cdot\mathbf{q}_e,
\end{equation}
\begin{equation}
  \label{eq:p-temp-eq}
   \frac{ \d T_p} { \d r} = -\frac{4}{3} \frac{T_p}{r}
                     + \frac{2}{3} \frac{1}{k_B n_p U} Q_p
                     + \frac{T_e - T_p}{U \tau},
\end{equation}
where $r$ is heliocentric distance,
$U$ is the bulk solar wind speed, $n_{\{e,p\}}$ is the electron (proton)
number density, $Q_{\{e,p\}}$ represents turbulent heating per unit mass,
$\mathbf{q}_e$ is the electron heat flux vector (proton heat flux
vector has been neglected), and
        $ k_B $ is Boltzmann's constant.
The $T_e - T_p$
terms in Eqs.~(\ref{eq:e-temp-eq})--(\ref{eq:p-temp-eq}) model
Coulomb collisions taking place over a
timescale       $\tau$  \citep[e.g.,][]{Priest}.

In previous models \citep{ZankEA96, MattEA99-swh, SmithEA01,
  BreechEA08} the equation for evolution of the temperature has been
supplemented by an equation for the turbulence energy, an equation for
the correlation or energy-containing scale and an equation for the
cross helicity.  These will be revisited further below.  Here we will
focus on issues surrounding the inclusion of the separate temperature
equations for protons and electrons.  In particular, three questions
arise by writing down these equations: (1) How does the electron heat
flux vary with distance?  (2) How much turbulent dissipation goes into
heating the electrons versus the protons? and (3) Over what timescale
do the electrons and protons experience energy exchange couplings,
such as Coulomb collisions, that tend to equilibrate their
temperatures?  

In this paper, we explore the physical issues surrounding these
questions within the context of a turbulence transport model for the
solar wind.  The physical issues may be further complicated by
  the effects of pickup protons and latitudinal variations present in
  the solar wind.  As a result, we limit ourselves to the high
  latitude fast wind, which displays less dependence on latitude, at
  least above 35 degrees or so \citep{McComasEA00}.  We also limit
  ourselves to examining these issues primarily in the inner parts of
  the heliosphere ($r < 10$ AU), which avoids complications due to
  pickup protons.  In future work, we will examine how pickup protons
  and latitudinal variations affect the key issues listed above.

\section{Electron Effects}
\label{sec:elect-effects}

\subsection{Electron Heat Flux}
\label{sec:electron-heat-flux}

One of the new effects we include is the electron
heat flux.   Generally speaking
the heat flux associated with the proton distribution
is regarded as unimportant \citep{Braginskii65}.   
For the case of the solar wind, a simple way to see this is to assume
that the proton heat flux
is due to a beam 
with fractional number density, $f_{\mathrm{beam}}$, 
moving at the Alfv\'en speed, $V_A$,
relative to
the bulk proton population.
This leads to a heat flux on the order of $f_{\mathrm{beam}} V_A^3$.
If we take the scale of its divergence to be $R$, the heliocentric
distance, then the relative
contribution
of the divergence of the proton heat flux
to the proton temperature given
in Eq.~(\ref{eq:p-temp-eq})
is of the order $f_{\mathrm{beam}}V_A^3/R$.
We compare this result to the heating term
        (see    \S\ref{sec:turb-heat} below),
which  is of order $ Z^3/\lambda $,   where $Z$ is
the fluctuation amplitude and $\lambda$ the correlation scale.
Since
$Z \sim V_A/2 $ and $\lambda \sim R/100$, while $f_{\mathrm{beam}}$ cannot be
greater than unity and is probably $< 1/10$, it is clear that
for protons the heat flux is much less important than the heating.
Electrons, on the other hand, with their  high thermal speed and
anisotropic distribution functions, are affected greatly by heat flux.
Within the solar wind, the electron heat flux is primarily   along the
magnetic field.   Heat flux transverse to the magnetic field is
severely reduced as the magnetic field effectively acts as  an  insulating
blanket.

Modeling $q_{_\parallel}$---the
        electron heat flux along the magnetic field---requires
approximating the solar wind as either
collision dominated or collisionless.   A
collision dominated
model,
$  q_{_\parallel} = -\kappa_{_\parallel} \nabla_{_\parallel} T_e (r)$
\citep{SpitzerHarm53} produces temperatures  that are  too large
at 1 AU compared to observations \citep{Hollweg76}.
Collisionless models \citep{Hollweg76}
fare much better and may be adequate for most purposes.
Nonetheless, the
collisionless
models can still suffer from missing non-local effects
\citep{CanulloEA96, ScudderOlbert79a}.

\begin{figure}
  \centering
  \includegraphics[scale=0.40]{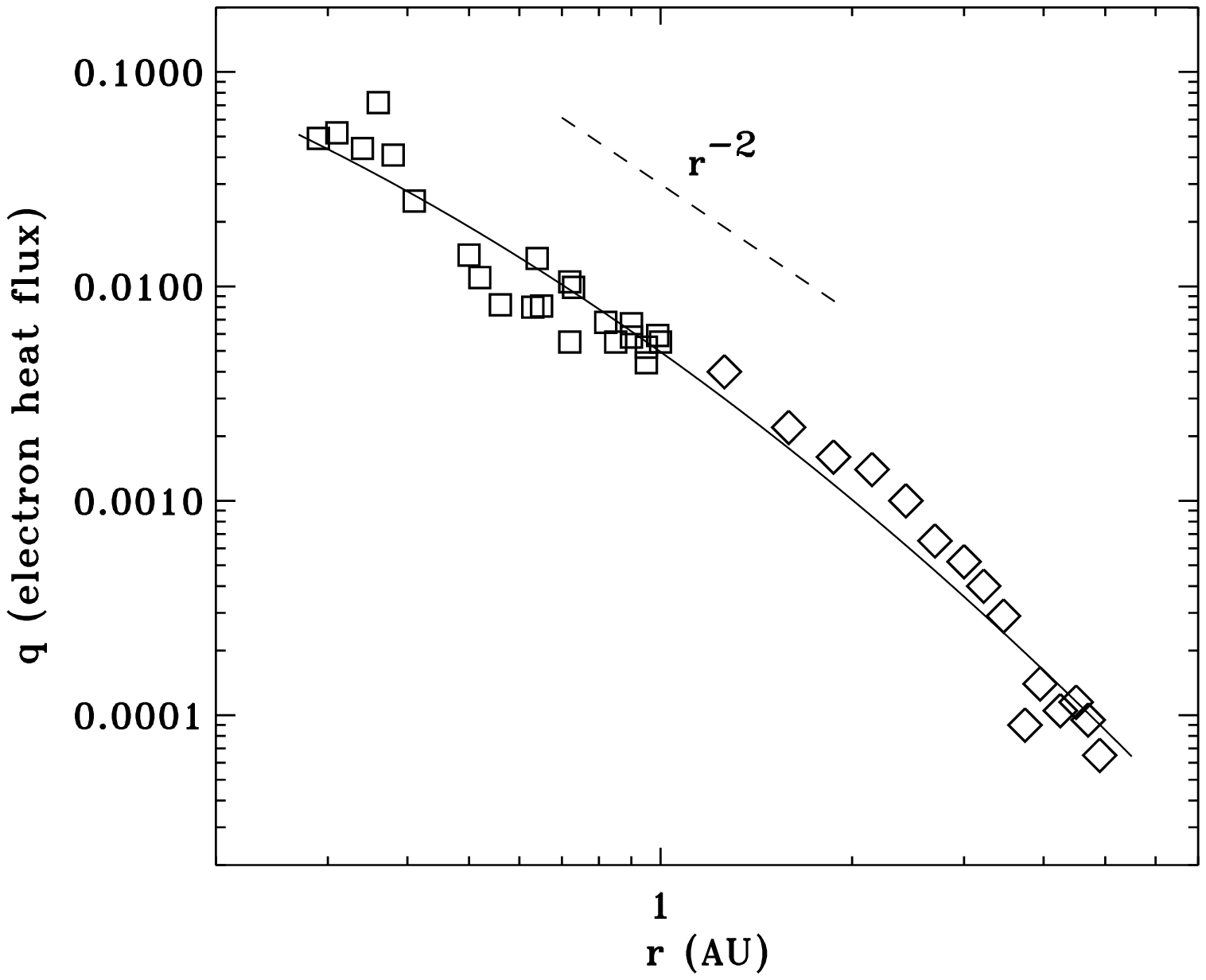}
  \caption{Observed electron heat flux (cgs units)
    \citep{PilippEA90, ScimeEA94}
    and fits to the data.}
  \label{fig:e-cond-data-fits}
\end{figure}

Rather than employing a theoretically based model,
here we adopt a more empirical approach in order to handle
these difficulties in specifying $q_{_\parallel}$.
We derive $q_{_\parallel}$ from
observational data by applying a best fit procedure to
the electron heat fluxes given by
\citet{PilippEA90} for  0.3\,AU to      1\,AU
and \citet{ScimeEA94} for
        1\,AU to 5.4\,AU.
 Figure~\ref{fig:e-cond-data-fits} shows the
observations and the fitted curve.
Introducing
  $ x \equiv \log_{10} (r / R_{\text{1\,AU}}) $,
the final fit is given by
\begin{eqnarray}
  \label{eq:model-q}
  \log_{10} q_{_\parallel} (r) = -2.3054 - 2.115 x - 0.58604 x^2 ,
\end{eqnarray}
and is assumed to be a reasonable approximation
out to $r \sim 100$\,AU.  
Note that past 10 AU, 
pickup proton effects may invalidate this assumption.  

We remark that the use of this electron heat conduction data
  may introduce some uncertainty into our results.  In particular, the
  measurements shown in Figure~\ref{fig:e-cond-data-fits} were adopted
  from \citet{ScimeEA94}.  The observations come from the initial
  Ulysses cruise phase from 1 to 5 AU, while Ulysses was in the
  ecliptic plane.  To our knowledge, no velocity selection was
  performed on the observations.  We therefore assume the measurements
  apply to all wind speed intervals and latitudes.  Note that in some
  later observations, \citet{ScimeEA99} found no significant variation
  in $q_{\parallel}$ at higher latitudes, which are dominated by the
  fast wind.  We will return to this point later when discussing
  our results.  

\subsection{Collision Effects}
\label{sec:coll-effects}

Protons and electrons can exchange heat through Coulomb collisions,
though they may do so over extremely long timescales.  They each can
also exchange energy with the electromagnetic field, and therefore
indirectly with each other, through wave-particle interactions.  These
kinetic effects may be thought of as producing effective collisions.
The observational fact that protons and electrons frequently maintain
different temperatures in the solar wind \citep{Hundhausen} provides
clear evidence that this coupling is not very strong.  
Generally, the scales for observing collisional effects
correspond to several AU, as has been found recently
\citep{KasperEA08} in a comparison of fast and slow wind
characteristics where proton-ion collisions are the
central consideration.


Taking into account this background,
we adopt a (somewhat) arbitrary approach for setting
the collision timescale, $\tau$.
For this work, we set $\tau$ as a constant
equal to
the plasma transit time to some distance (e.g., 10\,AU).
This choice produces only a weak interaction
between protons and electrons,
but does allow for collisional energy exchange effects to be seen in the model results.

\subsection{Partitioning Turbulent Heating}
\label{sec:turb-heat}

As the solar wind plasma evolves, MHD turbulence transfers energy from
low wavenumbers to high wavenumbers, where the energy is dissipated.
The dissipated energy heats both the protons and electrons.   Since we
are modeling the supply of energy from large scales to small scales, a
process believed to be controlled mainly by large-scale MHD processes,
we cannot, on the basis of this analysis, distinguish the channel or
sequence of kinetic processes that absorbs the energy.   Possibilities
are kinetic Alfv\'en waves
\citep{LeamonEA98b,CranmervanBall03,BaleEA05,GaryBorovsky08}, whistlers
\citep{GaryEA08}, and nonlinear dissipation in current sheets
\citep{SundkvistEA07}, to name a few.   To account for the data, however,
we will have to employ a reasonably correct partitioning of the
cascaded energy into the heat functions $Q_p$ and $Q_e$ that appear in
the temperature equations (\ref{eq:e-temp-eq}) and
(\ref{eq:p-temp-eq}).

Specifically,
let $Q$ be the total energy dissipated by the turbulence.   We
then define
\begin{equation}
  \label{eq:q-ep-def}
  Q_p \equiv f_p Q,
        \qquad
  Q_e \equiv (1 - f_p) Q ,
\end{equation}
where
$f_p$ is the fraction that determines the amount of
turbulent heating that goes into the
protons ($Q_p$).
The rest of the heat is given to the electrons ($Q_e$).
We make no distinction
here between the core, halo, and strahl populations of the electrons
as the turbulence heats all electrons.

In principle, $f_p$ can be determined by the underlying kinetic
physics \citep{GaryBorovsky04},
which operates  at scales far smaller than the energy-containing scale
described by the turbulence model.   $f_p$ may also change with
increasing distance 
  \citep[see the campanion paper]
        [for more details on how $f_p$ may change]{CranmerEA09}.   
We neglect these features of the
problem for now and simply adopt $f_p = \text{const}$.

Some evidence suggests
\citep{LeamonEA98b,LeamonEA99} that
        $f_p \approx  0.60$
at      1\,AU,
if one associates helicity-sensitive processes
with proton cyclotron absorption, and   the  remaining
heating as due to a process like Landau damping, which can
affect both protons and electrons.
There has also been the suggestion \citep{MattEA08-Taylor}
that a stronger cascade drives proton instabilities more robustly, leading
to more absorption near proton cyclotron scales.
However, lacking a firm quantitative theoretical basis,
such empirically based arguments must be regarded
with caution at present.  Therefore we
revert to a purely empirical choice of $f_p$, to be
supported or contraindicated by
 the results of the model, shown below.

For the total heating rate due to the cascade, $Q$,
we
employ  a
\citet{KarmanHowarth38} style phenomenology giving
\begin{equation}
  \label{eq:kh-style}
  Q \sim \frac{Z^3}{\lambda}
\end{equation}
with
$Z^2$ being (twice) the total energy available
to the turbulence and $\lambda$ being the length scale
of the largest turbulent structures.
Both of these quantities must be extracted from the turbulence
model.
Note that this dissipation model is rather robust and has
  been successfully 
  applied to very different regimes, such as
  the super Alfv\'enic solar wind \citep[e.g.,][]{BreechEA08}
  and the corona \citep[e.g.,][]{DmitrukEA01-apj, VerdiniVelli07}.

\section{Turbulence Model}

We adopt the following turbulence model
        \citep{ZankEA96,MattEA99-swh,MattEA04-Hc,
                SmithEA01,SmithEA06-pi,IsenbergEA03,
               BreechEA05,BreechEA08}
which comprises three equations describing the transport of
turbulence quantities; one equation for
(twice) the total turbulent energy, $Z^2$,
\begin{equation}
  \label{eq:model-Z2}
   \frac{ \d Z^2} { \d r}
  = -\frac{Z^2}{r}
    + \frac{\Csh - M\sigma_D}{r} Z^2
    + \frac{\dot{E}_{PI}}{U}
    - \alpha f^+ \frac{Z^3}{\lambda U} ,
\end{equation}
correlation (similarity) scale, $\lambda$,
\begin{equation}
  \label{eq:model-lambda}
   \frac{ \d \lambda} { \d r} = \beta f^+ \frac{Z}{U}
                  - \frac{\beta}{\alpha} \frac{\dot{E}_{PI}}{U Z^2} \lambda ,
\end{equation}
and the normalized cross helicity, $\sigma_c$,
\begin{equation}
  \label{eq:model-sc}
   \frac{ \d \sigma_c} { \d r}
  = \alpha f' \frac{Z }{U \lambda}
     -\left[\frac{\Csh - M\sigma_D}{r}
       +  \frac{\dot{E}_{PI}}{U Z^2}
      \right]\sigma_c ,
\end{equation}
where
\begin{equation}
        f^\pm(\sigma_c)
        =
        \frac{(1-\sigma_c^2)^{1/2}}{2}
        \left [
           (1+\sigma_c)^{1/2} \pm (1-\sigma_c)^{1/2}
        \right ]  ,
  \label{eq:fsigma}
\end{equation}
and
        $ f'(\sigma_c) = \sigma_c f^+ - f^-$.
Driving of the turbulence comes from shear, modeled through $\Csh$,
and pickup protons, $\dot{E}_{PI}$, with shear being more dominant
in the inner heliosphere (say, $r < 10 $\,AU)
and pickup protons more dominant in the outer heliosphere.
$M = 1/2$ relates to the underlying turbulence
geometries and
        $\sigma_{_D} = -1/3$ approximates the normalized energy
difference (kinetic minus magnetic)
of the fluctuations.
More details on the model itself, and its parameters,
can be found in \citet{BreechEA08}.

The rightmost term of Eq.~(\ref{eq:model-Z2}),
 $\alpha f^+ Z^3 / \lambda U$, controls the turbulent
dissipation.
The constant
        $\alpha $   (taken $= 2 \beta $ herein)
modulates the efficiency of
the dissipation, with higher values resulting in a more efficient
dissipation (e.g., more active turbulent cascade).
Previous use of the turbulent model
        \citep[such as][]{BreechEA08,SmithEA06-pi}
used values
of $\alpha = 0.8$.   For this work, we adopt lower values
of $\alpha = 0.5$ based on the results of
hydrodynamical simulations \citep{PearsonEA04}.   Even lower
values, down to $\alpha = 1/8$, may also be consistent with
those simulations.

The turbulence dissipates energy, which then goes into heating
of the solar wind and manifests as elevated electron and proton
temperatures.   The total turbulence heating rate
is given as

\begin{equation}
  \label{eq:turb-heating-def}
  Q = \frac{1}{2} \alpha \rho f^+ (\sigma_c) \frac{Z^3}{\lambda}
\end{equation}
which is partitioned between the electrons and protons
according to Eq.~(\ref{eq:q-ep-def}).
The factor of $1/2$ appears because $Z^2$ is twice the
energy density of the fluctuations.

\section{Model Results}
\label{sec:model-res}

\begin{figure}
  \centering
  \includegraphics[width=\columnwidth]{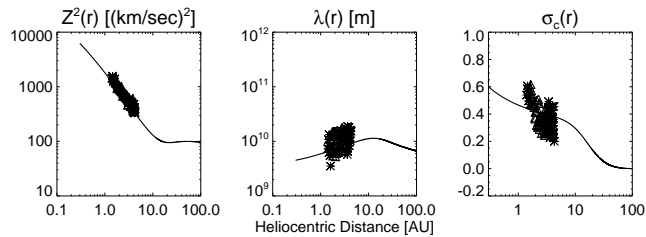}
  \caption{Solution of the turbulence model used in this paper.
       The boundary values are specified at  0.3\,AU
       as
        $Z^2=5000$\,(km/s)$^2$,
        $\lambda = 0.03$\,AU, and
        $\sigma_c = 0.6$.
       Other parameters are set
       as
        $U = 774$\,km/s,
        \Csh = 0.25,
        $\alpha = 2 \beta = 0.5$,
       and
        $M \sigma_D = -1/6$.
     Pickup protons effectively turn on near    10\,AU.
     The solutions (solid lines)
     compare reasonably well to the observed Ulysses
     values from near solar minimum \citep{BavassanoEA00a, BavassanoEA00b}.}
  \label{fig:turb-soln}
\end{figure}

We can numerically solve the transport model,
Eqs.~(\ref{eq:model-Z2})--(\ref{eq:model-sc}), and use
the results to compute the temperature values.
Figure~\ref{fig:turb-soln} shows the solutions of the turbulence
model.   The computed solutions compare well against
Ulysses observations of 1 hour cadence from solar minimum
        \citep{BavassanoEA00a,BavassanoEA00b}.
The model can also produce other solutions,
which bracket the observed values \citep[see][]{BreechEA08}.

The transport solutions can then be used to compute the proton and
electron temperatures using Eqs.~(\ref{eq:e-temp-eq})
and (\ref{eq:p-temp-eq})
and taking $f_p$ as a constant 0.6.   
The temperature solutions given below use
observations from the SWOOPS \citep{BameEA92} experiment
on Ulysses  corresponding to the same time as the
observations shown in Fig.~\ref{fig:turb-soln}.   
The temperature
  observations were made in the  high latitude, fast wind during
  solar minimum conditions.
For the proton
temperatures, we have used the geometric mean of the two temperature
values given in the data archives
\citep[see][for more information]{CranmerEA09}.   For
the electron temperatures,
SWOOPS provides temperature data for the core, halo, and total
electron populations.   We compare our results against the total
electron temperature as the turbulence heating should affect
all electrons.   For the model solutions, we take the initial
values of
        $ T_p = 2.0 \times 10^6 $\,K
and
        $ T_e = 4.0 \times 10^5 $\,K
at 0.3 AU.

\begin{figure}
  \centering
 \includegraphics[width=\columnwidth]{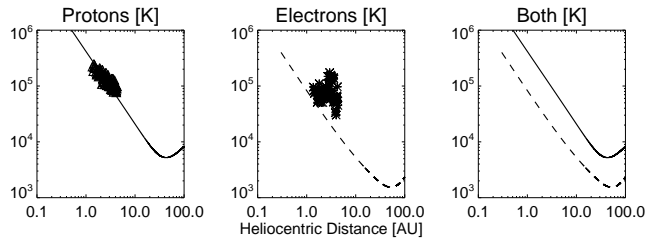}
  \caption{
    Model solutions for the proton and electron temperatures
            computed without collisions and without
            electron heat conduction.
            The proton solution matches the observations
            well, but the
            electron
            solution misses almost all of the  observed data.
          }
  \label{fig:tp-te-no-colls-divq}
\end{figure}

Figure~\ref{fig:tp-te-no-colls-divq} displays the temperature
solutions without  heat conduction and almost without
collisions.
The collision timescale $\tau$  was set equal
to the time for plasma to transit to 100\,AU,
which effectively removes collision effects from the results.
The proton solution agrees very well with the observed data,
but the electron solution misses almost all the data.

\begin{figure}
  \centering
  \includegraphics[width=\columnwidth]{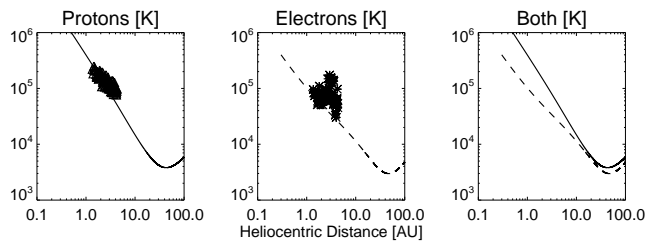}
  \caption{Model solutions for the proton and electron temperatures
            computed with collisions but without electron heat conduction.
            The proton solution is reasonable, but the
            electron
            solution falls below most of the observed data.
            Collisional effects are also apparent, with the two
            temperatures begining to equalize in the outer
            heliosphere, although the protons stay warmer.
          }
  \label{fig:tp-te-no-divq}
\end{figure}

Figure~\ref{fig:tp-te-no-divq} shows the temperature solutions without
the electron heat conduction, but with collisions turning on at 10\,AU
 (i.e., $\tau$ was set to the transit time to 10\,AU).
The proton solution mostly agrees with the
observations, but does show some disagreement near 5\,AU where the
proton temperature solution is slightly cooler than the observations
indicate.
Collisions evidently cause the slight disagreement as this
disagreement is not present
in Fig.~\ref{fig:tp-te-no-colls-divq}.   This may provide evidence that
setting the collision time to be the transit time to 10\,AU is too low
a value.

In contrast to the proton temperature solution, the electron solution
becomes slightly better when collisions are included.   The energy lost
from the protons profits the electrons.   Nonetheless, the electron
solution still misses   most of the observed data.

\begin{figure}
  \centering
  \includegraphics[width=\columnwidth]{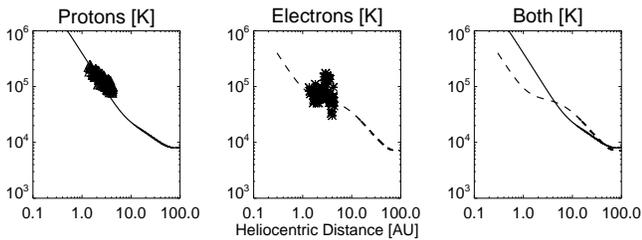}
  \caption{Model solutions for the proton and electron temperatures
            computed with both collisions and electron heat conduction
            turned on.
            The proton solution remains reasonable.
            The electron solution displays a ``shelf-like'' region
            between 1 and 10\,AU.   The solution matches the
            observations better than not including the
            heat conduction (cf.\ Fig.~\ref{fig:tp-te-no-divq}).
          }
  \label{fig:tp-te-all}
\end{figure}

Including both electron heat conduction, as determined
from equation~(\ref{eq:model-q}), and collisions produces
interesting results
        (Fig.~\ref{fig:tp-te-all}).
The electron temperature now features a ``shelf'' region between
1 and 10\,AU where $T_e(r)$  does not decrease as
rapidly
as it does outside that region.   Interestingly, the electron temperature
solution
crosses the proton temperature solution near 5\,AU, after which the
electrons are actually warmer than the protons.   Collisions then begin
to take hold,  resulting in higher proton temperatures and lower
electron
temperatures.

\begin{figure}
  \centering
 \includegraphics[scale=0.40]{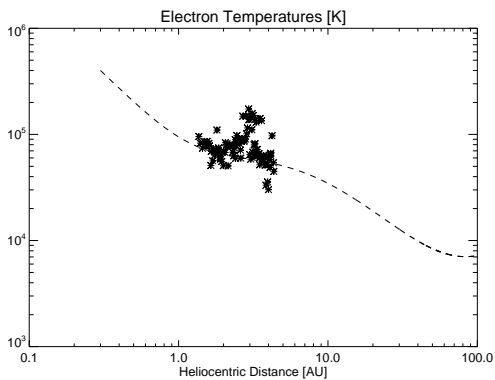}
  \caption{The electron temperature solution from
    Figure~\ref{fig:tp-te-all} showing the ``shelf''
    region more clearly.  The solution fits most of the data, although
    there are outliers, particularly around 3 AU.  The
    ``shelf'' region is due to heat conduction (see text).
          }
  \label{fig:te-only-data}
\end{figure}

Figure~\ref{fig:te-only-data} shows only 
the electron solution from Figure~\ref{fig:tp-te-all}.  
The shelf region matches the observed electron temperatures better
than
the solutions obtained when heat conduction is neglected.   While there is
some variability in the observations, the solution goes through most
of the available data.   Within this region, the heat conduction
evidently allows some heat to ``pile up'' to allow the electrons
to stay warmer than they may otherwise be.   The shelf  arises from
the electron heat flux shifting from a form close to $r^{-2}$ to
a steeper form in the outer heliosphere.

We note that the existence of the shelf region must be handled
carefully.  The shelf originates from the model solutions and compares
favorably with the data we have.  However, no high latitude data is
available beyond about 5 AU.  Further data would be required to
confirm the existence of shelf. Additionally, while we chose the
electron temperature data to match published data sets
\citep{BavassanoEA00a, BavassanoEA00b}, it may be possible that there
are data selection effects that lend to the appearance of the shelf.
The temperature data itself was taken solar minimum conditions and has
a cadence of 1 hour.  Selecting data with other cadences or different
solar cycle conditions could easily obscure the shelf \citep[see, for
instance,][]{PhillipsEA95c}.

\begin{figure}
  \centering
 \includegraphics[width=\columnwidth]{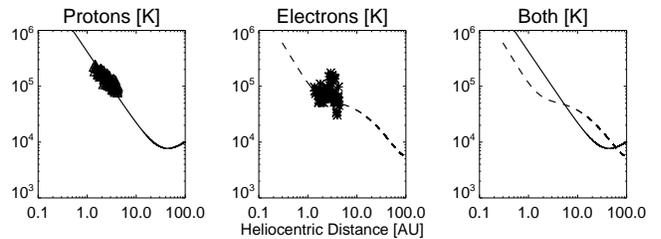}
  \caption{Model solutions for the proton and electron temperatures
            computed with electron heat conduction turned on, but
            without collisions.   The shelf in the electron
            temperature solution is still present, as is the
            crossover of the electron and proton solutions.
            The initial value of the electron temperature was
            raised to $6.0 \times 10^5 $\,K
            to maintain the good agreement with observations.
          }
  \label{fig:tp-te-divq-no-coll}
\end{figure}

The shelf is due solely to electron heat conduction, as
Fig.~\ref{fig:tp-te-divq-no-coll} reveals.
For the solutions shown there, collisions
were turned off
        (i.e., $\tau$ is set to the transit time to 100\,AU).
The shelf is still present, although the initial value of the
electron temperature had to be increased from
        $ 4.0 \times 10^5 $\,K
(as in the other solutions)
to $6.0 \times 10^5 $\,K
to maintain similar agreement with the
observations.   Note that the proton solution displays better agreement
with the observations due to the
lack of collisions.

\begin{figure}
  \centering
 \includegraphics[scale=0.40]{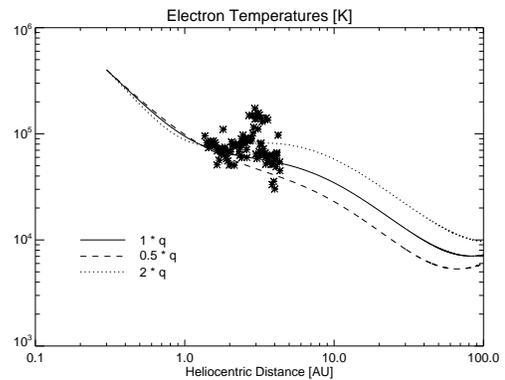}
  \caption{Solutions for the electron temperature using different
  profiles of the heat flux vector, $q_{\parallel}$.  The baseline
  solution uses the heat flux vector given in Eq.~(\ref{eq:model-q}).
  The other two solutions use baseline heat flux multiplied by
  2 and divided by 2.  Possible variations in the heat flux
  should be encompassed by those two extremes.  the figure
  demonstrates that a higher heat flux will make the shelf
  region more pronounced, while a lower heat flux lessens
  the shelf.
}
  \label{fig:q-bsoln}
\end{figure}

The shelf is somewhat sensitive to the underlying electron
  heat flux vector, $q_{\parallel}$, used to compute the solutions.
  Figure~\ref{fig:q-bsoln} shows the electron temperature solutions
  computed using three different profiles for $q_{\parallel}$.
  Equation~(\ref{eq:model-q}) provides the baseline profile,
  which yields the electron temperature solution shown
  in Figure~\ref{fig:te-only-data}).  The
  other two solutions use a heat flux vector where the baseline
  $q_{\parallel}$ was multiplied by 2 and divided by 2.  Since
  \citet{ScimeEA99} found small variations in the heat flux with
  latitude, any variation in the heat flux measurements should lie
  between these two extremes.  The electron temperature
  solutions show that increasing the
  heat flux makes the shelf more pronounced.  Lowering the heat flux
  lessens the shelf.  

\section{Conclusions}
\label{sec:conc}

For this work, we have added the effects of electrons to a turbulence
transport model for the solar wind.   Electron effects manifest in the
division of turbulent heating between protons and electrons,
collisions between protons and electrons, and heat conduction of the
electrons.   
We find that adding an empirical model for electron heat
conduction makes a dramatic difference in the solutions.   A
``shelf-like'' region appears between 1 and 10\,AU
where the electron
temperatures do not decrease as rapidly as they do outside that region.
The
electrons stay warm enough to actually become warmer than the protons.
Observations of the electron temperatures seem to support the
presence of the shelf.
We also find that collisions between the protons and electrons
may not be important until well past 10\,AU.
Allowing collisions to take effect closer in appears to
cause discrepancies between the model's proton temperature solution
and observed data.

%
%

\begin{acknowledgments}
This research was supported in part by an appointment to the
NASA Postdoctoral Program at Goddard Space Flight Center,
administered by Oak Ridge Associated Universities through
a contract with NASA.
SRC's work was supported by the National Aeronautics and Space
Administration (NASA) under grants NNG04GE77G, NNX06AG95G, and
NNX09AB27G to the Smithsonian Astrophysical Observatory.
WHM's work was supported by NSF ATM 0752135 (SHINE) and
NASA NNX08AI47G (Heliophysics Theory Program).
JCK's work was supported by NASA grant NNX08AW07G.
\end{acknowledgments}

\bibliographystyle{agu}
\bibliography{ag,hl,mp,qz,refs_bab}

\newcommand{\BIBand} {and} 
  \newcommand{\boldVol}[1] {\textbf{#1}} 
  \newcommand{\SortNoop}[1] {} 
  \newcommand{\au} {{A}{U}\ } 
  \newcommand{\AU} {{A}{U}\ } 
  \newcommand{\MHD} {{M}{H}{D}\ } 
  \newcommand{\mhd} {{M}{H}{D}\ } 
  \newcommand{\RMHD} {{R}{M}{H}{D}\ } 
  \newcommand{\rmhd} {{R}{M}{H}{D}\ } 
  \newcommand{\wkb} {{W}{K}{B}\ } 
  \newcommand{\alfven} {{A}lfv\'en\ } 
  \newcommand{\Alfven} {{A}lfv\'en\ } 
  \newcommand{\alfvenic} {{A}lfv\'enic\ } 
  \newcommand{\Alfvenic} {{A}lfv\'enic\ }
\begin{thebibliography}{45}
\providecommand{\natexlab}[1]{#1}
\expandafter\ifx\csname urlstyle\endcsname\relax
  \providecommand{\doi}[1]{doi:\discretionary{}{}{}#1}\else
  \providecommand{\doi}{doi:\discretionary{}{}{}\begingroup
  \urlstyle{rm}\Url}\fi

\bibitem[{\textit{Bale et~al.}(2005)\textit{Bale, Kellogg, Mozer, Horbury, and
  Reme}}]{BaleEA05}
Bale, S.~D., P.~J. Kellogg, F.~S. Mozer, T.~S. Horbury, and H.~Reme,
  Measurement of the electric fluctuation spectrum of magnetohydrodynamic
  turbulence, \textit{Phys.\ Rev.\ Lett.}, \textit{94}, 215002,
  \doi{10.1103/PhysRevLett.94.215002}, 2005.

\bibitem[{\textit{Bame et~al.}(1992)\textit{Bame, McComas, Barraclough, Sofaly,
  Chavez, Goldstein, and Sakurai}}]{BameEA92}
Bame, S., D.~McComas, B.~Barraclough, K.~Sofaly, J.~Chavez, B.~Goldstein, and
  R.~Sakurai, The {Ulysses} solar wind plasma experiment, \textit{Astron.\
  Astrophys. Supp.}, \textit{92}, 237--265, 1992.

\bibitem[{\textit{Bavassano et~al.}(2000{\natexlab{a}})\textit{Bavassano,
  Pietropaolo, and Bruno}}]{BavassanoEA00a}
Bavassano, B., E.~Pietropaolo, and R.~Bruno, \alfvenic turbulence in the polar
  wind: {A} statistical study on cross helicity and residual energy variations,
  \textit{J.\ Geophys.\ Res.}, \textit{105}, 12\,697--12\,704,
  2000{\natexlab{a}}.

\bibitem[{\textit{Bavassano et~al.}(2000{\natexlab{b}})\textit{Bavassano,
  Pietropaolo, and Bruno}}]{BavassanoEA00b}
Bavassano, B., E.~Pietropaolo, and R.~Bruno, On the evolution of outward and
  inward \alfvenic fluctuations in the polar wind, \textit{J.\ Geophys.\ Res.},
  \textit{105}, 15\,959--15\,964, 2000{\natexlab{b}}.

\bibitem[{\textit{Braginskii}(1965)}]{Braginskii65}
Braginskii, S.~I., Transport processes in a plasma, \textit{Rev. Plasma Phys.},
  \textit{1}, 205, 1965.

\bibitem[{\textit{Breech et~al.}(2005)\textit{Breech, Matthaeus, Minnie,
  Oughton, Parhi, Bieber, and Bavassano}}]{BreechEA05}
Breech, B., W.~H. Matthaeus, J.~Minnie, S.~Oughton, S.~Parhi, J.~W. Bieber, and
  B.~Bavassano, Radial evolution of cross helicity in high-latitude solar wind,
  \textit{Geophys.\ Res.\ Lett.}, \textit{32}, L06103,
  \doi{10.1029/2004GL022321}, 2005.

\bibitem[{\textit{Breech et~al.}(2008)\textit{Breech, Matthaeus, Minnie,
  Bieber, Oughton, Smith, and Isenberg}}]{BreechEA08}
Breech, B., W.~H. Matthaeus, J.~Minnie, J.~W. Bieber, S.~Oughton, C.~W. Smith,
  and P.~A. Isenberg, Turbulence transport throughout the heliosphere,
  \textit{J.\ Geophys.\ Res.}, \textit{113}, A08105,
  \doi{10.1029/2007JA012711}, 2008.

\bibitem[{\textit{Canullo et~al.}(1996)\textit{Canullo, Costa, and
  Ferro-{Fontan}}}]{CanulloEA96}
Canullo, M.~V., A.~Costa, and C.~Ferro-{Fontan}, Nonlocal heat transport in the
  solar wind, \textit{Astrophys.\ J.}, \textit{462}, 1005--1010, 1996.

\bibitem[{\textit{Coleman}(1968)}]{Coleman68}
Coleman, P.~J., Turbulence, viscosity, and dissipation in the solar wind
  plasma, \textit{Astrophys.\ J.}, \textit{153}, 371--388, 1968.

\bibitem[{\textit{Cranmer et~al.}(2009)\textit{Cranmer, Matthaeus, Breech, and
  Kasper}}]{CranmerEA09}
Cranmer, S., W.~Matthaeus, B.~Breech, and J.~Kasper, Empirical constraints on
  proton and electron heating in the inner heliosphere,
  \textit{\textsl{submitted to ApJ}}, 2009.

\bibitem[{\textit{Cranmer and {van}~{Ballegooijen}}(2003)}]{CranmervanBall03}
Cranmer, S.~R., and A.~A. {van}~{Ballegooijen}, \alfvenic turbulence in the
  extended solar corona: {Kinetic} effects and proton heating,
  \textit{Astrophys.\ J.}, \textit{594}, 573--591, 2003.

\bibitem[{\textit{Cranmer et~al.}(2007)\textit{Cranmer, {van}~{Ballegooijen},
  and Edgar}}]{CranmerEA07}
Cranmer, S.~R., A.~A. {van}~{Ballegooijen}, and R.~Edgar, Self-consistent
  coronal heating and solar wind acceleration from anisotropic
  magnetohydrodynamic turbulence, \textit{Astrophys.\ J.\ Suppl. Ser},
  \textit{171}, 520--551, \doi{10.1086/518001}, 2007.

\bibitem[{\textit{Dmitruk et~al.}(2001)\textit{Dmitruk, Milano, and
  Matthaeus}}]{DmitrukEA01-apj}
Dmitruk, P., L.~J. Milano, and W.~H. Matthaeus, Wave-driven turbulent coronal
  heating in open field line regions: {Nonlinear} phenomenological model,
  \textit{Astrophys.\ J.}, \textit{548}, 482--491, 2001.

\bibitem[{\textit{{Gary} and {Borovsky}}(2004)}]{GaryBorovsky04}
{Gary}, S.~P., and J.~E. {Borovsky}, {Alfv{\'e}n-cyclotron fluctuations: Linear
  Vlasov theory}, \textit{J.\ Geophys.\ Res.}, \textit{109}(A18), 6105,
  \doi{10.1029/2004JA010399}, 2004.

\bibitem[{\textit{{Gary} and {Borovsky}}(2008)}]{GaryBorovsky08}
{Gary}, S.~P., and J.~E. {Borovsky}, {Damping of long-wavelength kinetic
  Alfv{\'e}n fluctuations: Linear theory}, \textit{J.\ Geophys.\ Res.},
  \textit{113}(A12), 12,104, \doi{10.1029/2008JA013565}, 2008.

\bibitem[{\textit{{Gary} et~al.}(2008)\textit{{Gary}, {Saito}, and
  {Li}}}]{GaryEA08}
{Gary}, S.~P., S.~{Saito}, and H.~{Li}, {Cascade of whistler turbulence:
  Particle-in-cell simulations}, \textit{Geophys.\ Res.\ Lett.}, \textit{35},
  2104, \doi{10.1029/2007GL032327}, 2008.

\bibitem[{\textit{Hollweg}(1976)}]{Hollweg76}
Hollweg, J.~V., Collisionless electron heat conduction in the solar wind,
  \textit{J.\ Geophys.\ Res.}, \textit{81}, 1649--1658, 1976.

\bibitem[{\textit{Hundhausen}(1972)}]{Hundhausen}
Hundhausen, A.~J., \textit{Coronal Expansion and the Solar Wind},
  Springer-Verlag, New York, 1972.

\bibitem[{\textit{Isenberg et~al.}(2003)\textit{Isenberg, Smith, and
  Matthaeus}}]{IsenbergEA03}
Isenberg, P.~A., C.~W. Smith, and W.~H. Matthaeus, Turbulent heating of the
  distant solar wind by interstellar pickup protons, \textit{Astrophys.\ J.},
  \textit{592}, 564--573, 2003.

\bibitem[{\textit{Kasper et~al.}(2008)\textit{Kasper, Lazarus, and
  Gary}}]{KasperEA08}
Kasper, J.~C., A.~J. Lazarus, and S.~P. Gary, Hot solar-wind {Helium}: {Direct}
  evidence for local heating by \alfven-{Cyclotron} dissipation, \textit{Phys.\
  Rev.\ Lett.}, \textit{101}(26), 261103, \doi{10.1103/PhysRevLett.101.261103},
  2008.

\bibitem[{\textit{Leamon et~al.}(1998)\textit{Leamon, Matthaeus, Smith, and
  Wong}}]{LeamonEA98b}
Leamon, R.~J., W.~H. Matthaeus, C.~W. Smith, and H.~K. Wong, Contribution of
  cyclotron-resonant damping to kinetic dissipation of interplanetary
  turbulence, \textit{Astrophys.\ J.}, \textit{507}, L181, 1998.

\bibitem[{\textit{Leamon et~al.}(1999)\textit{Leamon, Smith, Ness, and
  Wong}}]{LeamonEA99}
Leamon, R.~J., C.~W. Smith, N.~F. Ness, and H.~K. Wong, Dissipation range
  dynamics: {K}inetic \alfven waves and the importance of $\beta_e$,
  \textit{J.\ Geophys.\ Res.}, \textit{104}, 22\,331, 1999.

\bibitem[{\textit{Mac{Bride} et~al.}(2008)\textit{Mac{Bride}, Smith, and
  Forman}}]{MacBrideEA08}
Mac{Bride}, B.~T., C.~W. Smith, and M.~A. Forman, The turbulent cascade at
  1\,{A}{U}: {Energy} transfer and the third-order scaling for {M}{H}{D},
  \textit{Astrophys.\ J.}, \textit{679}(2), 1644--1660, \doi{10.1086/529575},
  2008.

\bibitem[{\textit{Marino et~al.}(2008)\textit{Marino, {Sorriso-Valvo}, Carbone,
  Noullez, Bruno, and Bavassano}}]{MarinoEA08}
Marino, R., L.~{Sorriso-Valvo}, V.~Carbone, A.~Noullez, R.~Bruno, and
  B.~Bavassano, Heating the solar wind by a magnetohydrodynamic turbulent
  energy cascade, \textit{Astrophys.\ J.}, \textit{677}, L71--L74,
  \doi{10.1086/587957}, 2008.

\bibitem[{\textit{Marsch and Tu}(1993)}]{MarschTu93c}
Marsch, E., and C.-Y. Tu, Modeling results on spatial transport and spectral
  transfer of solar wind \alfvenic turbulence, \textit{J.\ Geophys.\ Res.},
  \textit{98}, 21\,045, 1993.

\bibitem[{\textit{Matthaeus et~al.}(1999)\textit{Matthaeus, Zank, Smith, and
  Oughton}}]{MattEA99-swh}
Matthaeus, W.~H., G.~P. Zank, C.~W. Smith, and S.~Oughton, Turbulence, spatial
  transport, and heating of the solar wind, \textit{Phys.\ Rev.\ Lett.},
  \textit{82}, 3444--3447, \doi{10.1103/PhysRevLett.82.3444}, 1999.

\bibitem[{\textit{Matthaeus et~al.}(2004)\textit{Matthaeus, Minnie, Breech,
  Parhi, Bieber, and Oughton}}]{MattEA04-Hc}
Matthaeus, W.~H., J.~Minnie, B.~Breech, S.~Parhi, J.~W. Bieber, and S.~Oughton,
  Transport of cross helicity and the radial evolution of {Alfv\'enicity} in
  the solar wind, \textit{Geophys.\ Res.\ Lett.}, \textit{31}, L12803,
  \doi{10.1029/2004GL019645}, 2004.

\bibitem[{\textit{{Matthaeus} et~al.}(2008)\textit{{Matthaeus}, {Weygand},
  {Chuychai}, {Dasso}, {Smith}, and {Kivelson}}}]{MattEA08-Taylor}
{Matthaeus}, W.~H., J.~M. {Weygand}, P.~{Chuychai}, S.~{Dasso}, C.~W. {Smith},
  and M.~G. {Kivelson}, Interplanetary magnetic {Taylor} microscale and
  implications for plasma dissipation, \textit{Astrophys.\ J.}, \textit{678},
  L141--L144, \doi{10.1086/588525}, 2008.

\bibitem[{\textit{Mc{C}omas et~al.}(2000)}]{McComasEA00}
Mc{C}omas, D.~J., et~al., Solar wind observations over {Ulysses}' first full
  polar orbit, \textit{J.\ Geophys.\ Res.}, \textit{105}, 10\,419--10\,434,
  2000.

\bibitem[{\textit{Oughton and Matthaeus}(1995)}]{OughtonMatt95}
Oughton, S., and W.~H. Matthaeus, Linear transport of solar wind fluctuations,
  \textit{J.\ Geophys.\ Res.}, \textit{100}, 14\,783--14\,799, 1995.

\bibitem[{\textit{Pearson et~al.}(2004)\textit{Pearson, Yousef, Haugen,
  Brandenburg, and Krogstad}}]{PearsonEA04}
Pearson, B.~R., T.~A. Yousef, N.~E.~L. Haugen, A.~Brandenburg, and P.-A.
  Krogstad, Delayed correlation between turbulent energy injection and
  dissipation, \textit{Phys.\ Rev.\ E}, \textit{70}, 056301,
  \doi{10.1103/PhysRevE.70.056301}, 2004.

\bibitem[{\textit{Phillips et~al.}(1995)\textit{Phillips, Bame, Gary, Gosling,
  Scime, and Forsyth}}]{PhillipsEA95c}
Phillips, J.~L., S.~J. Bame, S.~P. Gary, T.~Gosling, E.~E. Scime, and R.~J.
  Forsyth, Radial and meridional trends in solar wind thermal electron
  temperature and anisotropy: {U}lysses, \textit{Space Sci.\ Rev.},
  \textit{72}, 109, 1995.

\bibitem[{\textit{Pilipp et~al.}(1990)\textit{Pilipp, Miggenrieder,
  M\"uhlh\"auser, Rosenbauer, and Schwenn}}]{PilippEA90}
Pilipp, W., H.~Miggenrieder, K.~M\"uhlh\"auser, H.~Rosenbauer, and R.~Schwenn,
  Large-scale variations of thermal electron parameters in the solar wind
  between 0.3 and 1\,{A}{U}, \textit{J.\ Geophys.\ Res.}, \textit{95},
  6305--6329, 1990.

\bibitem[{\textit{Priest}(1982)}]{Priest}
Priest, E.~R., \textit{Solar Magnetohydrodynamics}, Reidel, Dordrecht, Holland,
  1982.

\bibitem[{\textit{Scime et~al.}(1994)\textit{Scime, Bame, Feldman, Gary,
  Phillips, and Balogh}}]{ScimeEA94}
Scime, E.~E., S.~J. Bame, W.~C. Feldman, S.~P. Gary, J.~L. Phillips, and
  A.~Balogh, Regulation of the solar wind electron heat flux from 1 to 5 {AU}:
  {U}lysses observations, \textit{J.\ Geophys.\ Res.}, \textit{99},
  23,401--23,410, 1994.

\bibitem[{\textit{Scime et~al.}(1999)\textit{Scime, Jr, and
  Littleton}}]{ScimeEA99}
Scime, E.~E., A.~E.~B. Jr, and J.~Littleton, The electron heat flux in the
  polar solar wind: {Ulysses} observations, \textit{Geophys.\ Res.\ Lett.},
  \textit{26}(14), 2129--2132, 1999.

\bibitem[{\textit{Scudder and Olbert}(1979)}]{ScudderOlbert79a}
Scudder, J.~D., and S.~Olbert, {A Theory of Local and Global Processes which
  affect Solar Wind Electrons 1. The Origin of Typical 1 AU Velocity
  Distribution Functions---Steady State Theory}, \textit{J.\ Geophys.\ Res.},
  \textit{84}, 2755--2772, 1979.

\bibitem[{\textit{Smith et~al.}(2001)\textit{Smith, Matthaeus, Zank, Ness,
  Oughton, and Richardson}}]{SmithEA01}
Smith, C.~W., W.~H. Matthaeus, G.~P. Zank, N.~F. Ness, S.~Oughton, and J.~D.
  Richardson, Heating of the low-latitude solar wind by dissipation of
  turbulent magnetic fluctuations, \textit{J.\ Geophys.\ Res.}, \textit{106},
  8253--8272, 2001.

\bibitem[{\textit{Smith et~al.}(2006)\textit{Smith, Isenberg, Matthaeus, and
  Richardson}}]{SmithEA06-pi}
Smith, C.~W., P.~A. Isenberg, W.~H. Matthaeus, and J.~D. Richardson, Turbulent
  heating of the solar wind by newborn interstellar pickup protons,
  \textit{Astrophys.\ J.}, \textit{638}, 508--517, 2006.

\bibitem[{\textit{Spitzer and H\"arm}(1953)}]{SpitzerHarm53}
Spitzer, L., and R.~H\"arm, Transport phenomena in a completely ionized gas,
  \textit{Phys.\ Rev.}, \textit{89}(5), 977--981, 1953.

\bibitem[{\textit{Sundkvist et~al.}(2007)\textit{Sundkvist, Retino, Vaivads,
  and Bale}}]{SundkvistEA07}
Sundkvist, D., A.~Retino, A.~Vaivads, and S.~D. Bale, Dissipation in turbulent
  plasma due to reconnection in thin current sheets, \textit{Phys.\ Rev.\
  Lett.}, \textit{99}, 025004, \doi{10.1103/PhysRevLett.99.025004}, 2007.

\bibitem[{\textit{Verdini and Velli}(2007)}]{VerdiniVelli07}
Verdini, A., and M.~Velli, \alfven waves and turbulence in the solar atmosphere
  and solar wind, \textit{Astrophys.\ J.}, \textit{662}, 669--676, 2007.

\bibitem[{\textit{von K\'arm\'an and Howarth}(1938)}]{KarmanHowarth38}
von K\'arm\'an, T., and L.~Howarth, On the statistical theory of isotropic
  turbulence, \textit{Proc. Roy. Soc. London Ser. A}, \textit{164}, 192--215,
  1938.

\bibitem[{\textit{Zank et~al.}(1996)\textit{Zank, Matthaeus, and
  Smith}}]{ZankEA96}
Zank, G.~P., W.~H. Matthaeus, and C.~W. Smith, Evolution of turbulent magnetic
  fluctuation power with heliocentric distance, \textit{J.\ Geophys.\ Res.},
  \textit{101}, 17\,093, 1996.

\bibitem[{\textit{Zhou and Matthaeus}(1990)}]{ZhouMatt90a}
Zhou, Y., and W.~H. Matthaeus, Transport and turbulence modeling of solar wind
  fluctuations, \textit{J.\ Geophys.\ Res.}, \textit{95}, 10\,291, 1990.

\end{thebibliography}

\end{article}

\end{document}